\documentclass[preprint,12pt,authoryear]{elsarticle}
\biboptions{sort}
\usepackage{amsmath,amssymb,amsfonts}
\usepackage{graphicx}
\usepackage{etoolbox}
\usepackage{iftex}
\ifPDFTeX\pdfminorversion=7\fi
\usepackage{xcolor} % 

\newcommand{\dsyrev}[1]{\textcolor{black}{#1}}

\usepackage{hyperref}
\hypersetup{hidelinks}
\usepackage{textcomp}
\usepackage{booktabs}
\usepackage{multirow}
\usepackage[pass]{geometry}
\usepackage{pdflscape}
\usepackage{placeins}
\def\BibTeX{{\rm B\kern-.05em{\sc i\kern-.025em b}\kern-.08em
    T\kern-.1667em\lower.7ex\hbox{E}\kern-.125emX}}
\journal{Computerized Medical Imaging and Graphics}
\begin{document}
\begin{frontmatter}
\title{Neurodegenerative Differential Diagnosis with Knowledge-enhanced Structural Reasoning}
\author[aff1]{Mingxuan Wang\fnref{equal}}
\ead{wmx0606@sjtu.edu.cn}
\author[aff2]{Siyuan Du\fnref{equal}}
\author[aff3]{Yanfeng Wang}
\author[aff3]{Ya Zhang}
\author[aff4]{Chencheng Zhang}
\author[aff5,aff3]{Jiangchao Yao\corref{cor1}}
\ead{sunarker@sjtu.edu.cn}
\address[aff1]{SJTU Paris Elite Institute of Technology, Shanghai Jiao Tong University}
\address[aff2]{School of Computer Science, Fudan University}
\address[aff3]{School of Artificial Intelligence, Shanghai Jiao Tong University}
\address[aff4]{Ruijin Hospital, Shanghai Jiao Tong University}
\address[aff5]{Cooperative Medianet Innovation Center (CMIC)}
\fntext[equal]{Mingxuan Wang and Siyuan Du contributed equally to this work.}
\cortext[cor1]{Corresponding author: J. Yao (sunarker@sjtu.edu.cn).}

\begin{abstract}
Accurate differential diagnosis of neurodegenerative disorders remains challenging due to overlapping clinical manifestations and heterogeneous, often incomplete multimodal observations. Existing multimodal models can integrate imaging and clinical measurements but often overlook disease-specific evidence patterns and pairwise relations among competing diagnoses. We therefore propose Knowledge-enhanced Structural Reasoning (KnowSR), which aligns multimodal patient representations with a literature-grounded knowledge bank of disease prototypes, signed evidence directions, and pairwise diagnostic relations. Specifically, in KnowSR, patient-specific knowledge alignment and target-aware pooling inject structured disease evidence, while anchored relational reasoning reconciles pairwise diagnostic margins over a sparse differential graph, thereby collectively enhancing differential diagnostic capability. Across extensive experiments, KnowSR consistently outperforms 19 baselines on patient-level six-class diagnosis under patient-disjoint five-fold evaluation. This shows that grounding patient evidence in structured medical knowledge enables more effective differential diagnosis than naive multimodal fusion alone.

\end{abstract}
\end{frontmatter}

%三点一行+大致方法
%第一段：背景，事情 第二段：现有做法+看法 第三段：哪些事情需要解决+自己的想法 第四段：自己方法
\section{Introduction}
\label{sec:introduction}

Neurodegenerative disorders often exhibit overlapping cognitive, behavioral, and motor manifestations despite distinct underlying pathologies. Distinguishing Alzheimer's disease (AD), Parkinson's disease (PD), frontotemporal dementia (FTD), progressive supranuclear palsy (PSP), and corticobasal syndrome (CBS) from normal cognition (NC) and from one another therefore requires integrating clinical, imaging, and laboratory evidence~\citep{jack2024revised,postuma2015mds,rascovsky2011bvftd,hoglinger2017psp,armstrong2013cbd}.
% as reflected in contemporary diagnostic criteria
%这里可以提前布局一下我们的central challenge
\dsyrev{The key is to determine how the available evidence favors one diagnosis over clinically similar alternatives.}
Computationally, this setting is naturally formulated as differential diagnosis under heterogeneous and incomplete multimodal evidence.

Existing computational approaches to neurodegenerative diagnosis have extensively exploited resting-state functional magnetic resonance imaging (rs-fMRI), which provides a noninvasive window into distributed brain dynamics, and graph- and transformer-based models have substantially improved functional representation learning~\citep{li2021braingnn,kan2022brainnetworktransformer,zhang2023lggnn}.
Because neurodegenerative disorders are multifactorial and cannot be fully characterized by imaging alone, recent multimodal approaches further integrate clinical scales, demographics, genetic factors, and fluid biomarkers~\citep{qiu2022multimodal,golovanevsky2022multimodal,yu2024transformer}, while incomplete observations are addressed through imputation, modality dropout, or specialized fusion architectures~\citep{havaei2016hemis,yu2024smart,han2024fusemoe,Du2026DementiaED}. 
Despite these advances, most existing systems often optimize latent representations for classification without explicitly assessing whether patient findings support specific diseases, conform to expected clinical directions, or distinguish among competing diagnoses.
Medical knowledge has been incorporated for representation initialization, retrieval, and post hoc explanation~\citep{chandak2023primekg,singhal2023clinical,thirunavukarasu2023llm,selvaraju2017gradcam,lundberg2017unified}, \dsyrev{but is rarely used to structure diagnostic reasoning.}

% As illustrated in Fig.~\ref{fig:motivation}, the central challenge is to distinguish clinically similar neurodegenerative disorders despite substantial overlap in cognitive, behavioral, and motor manifestations. 
\dsyrev{The central challenge is to distinguish clinically similar neurodegenerative diagnoses from incomplete patient observations when clinical manifestations overlap.}
%
% Achieving this objective under real-world multimodal settings involves two inherent challenges. 
%进一步解释挑战
\dsyrev{A finding consistent with one disease may also occur in another, so disease compatibility alone does not resolve diagnostic ambiguity.}
Missing measurements further restrict the evidence available for these comparisons, requiring informative signals to be retained without introducing artifacts from missing data (\textit{evidence preservation}).
%加了一点逻辑，这俩感觉并不像并列逻辑
% \dsyrev{Yet retaining an observed finding does not establish its diagnostic value: the same finding may support several competing diseases.}
%
\dsyrev{The available evidence must therefore be linked to the diagnoses it supports or contradicts and to the alternatives it helps distinguish (\textit{evidence interpretability}). }

\begin{figure}[!htbp]
    \centering
    \includegraphics[width=\columnwidth]{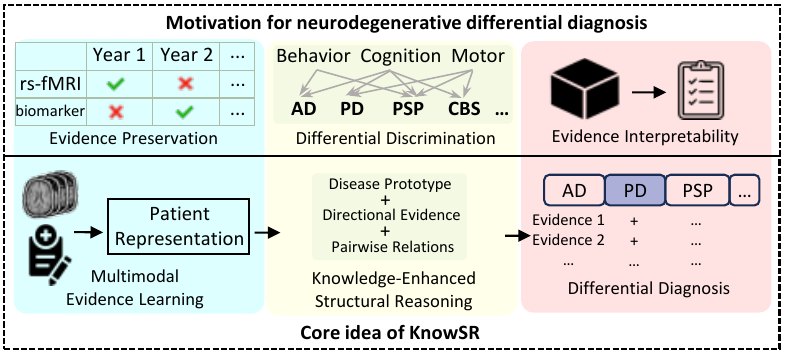}
    \caption{Motivation and core idea of KnowSR for neurodegenerative differential diagnosis. The upper row illustrates the central discrimination problem arising from overlapping clinical manifestations under incomplete multimodal observations, with the requirement of preserving available evidence and providing interpretable diagnostic support. The lower row summarizes the KnowSR framework.}
    \label{fig:motivation}
\end{figure}

\dsyrev{To resolve this ambiguity, our central idea is to organize patient evidence at two complementary levels: support for individual diagnoses and relative support between competing diagnoses (Fig.~\ref{fig:motivation}).}
\dsyrev{Disease prototypes and directional evidence establish a patient-specific diagnostic anchor, which is then refined by reconciling pairwise evidence across clinically related diagnoses.}
We \dsyrev{therefore} propose Knowledge-enhanced Structural Reasoning (KnowSR), which combines multimodal patient representations with structured disease-specific and relational knowledge for evidence-grounded neurodegenerative differential diagnosis.
The main contributions of this work are summarized as follows:
\begin{itemize}
%这里感觉直接说重点比较好
\item We 
% formulate neurodegenerative differential diagnosis as a unified six-class task and
develop an evidence-preserving multimodal framework \dsyrev{for unified six-class neurodegenerative differential diagnosis.}
It retains observed information while excluding unavailable measurements from cross-modal interaction.

%这里感觉技术概念一句话里面太多了，分成两句吧。
\item We introduce a literature-grounded knowledge bank comprising disease prototypes, directional evidence, and pairwise relations. \dsyrev{Patient-specific evidence alignment and anchored relational reasoning integrate this knowledge to distinguish among competing diagnoses.}

%最后一条贡献应该更像“发现了什么”，而不是“测了什么”。以及尽量内容量跟前面两点差不多。
% \item We evaluate KnowSR using patient-disjoint five-fold validation against 19 baselines. Database-specific, ablation, and evidence analyses demonstrate improved six-class diagnosis and assess cross-modal interaction, structured knowledge, relational refinement, Card semantics, and anatomical attention grounding.
\item \dsyrev{We demonstrate improved six-class diagnostic performance over 19 baselines under patient-disjoint five-fold evaluation. Ablation and evidence analyses confirm the value of structured knowledge and relational refinement for evidence-grounded differential diagnosis.}

\end{itemize}
\section{Related Work}
\label{sec:related_work}

\subsection{Data-driven and Multimodal Neurodegenerative Diagnosis}

Early computational studies of neurodegenerative diagnosis commonly relied on conventional classifiers, such as support vector machines, logistic regression, random forests, and gradient-boosted trees, using handcrafted imaging or clinical features~\citep{cortes1995support,breiman2001random,chen2016xgboost}. These approaches provide useful references for evaluating learned representations, but their performance largely depends on predefined feature extraction. More recent deep models employ recurrent networks and Transformer-based architectures to learn nonlinear temporal and cross-modal interactions directly from heterogeneous observations~\citep{schuster1997bidirectional,tsai2019multimodal}.

Multimodal learning has further become an important direction for neurodegenerative diagnosis. Qiu et al.~\citeyearpar{qiu2022multimodal} showed that combining neuroimaging with complementary clinical information can improve dementia-related classification, while Xue et al.~\citeyearpar{xue2024ai} further demonstrated the value of integrating heterogeneous clinical and biomarker evidence. Methods such as MedFuse and DrFuse additionally address incomplete multimodal observations through flexible and disentangled fusion strategies~\citep{hayat2022medfuse,yao2024drfuse}. Despite these advances, most existing approaches primarily optimize predictive representations from the available data and do not explicitly organize patient evidence according to competing diagnostic hypotheses.

\subsection{Brain Representation Learning and Knowledge Guidance}

Resting-state functional magnetic resonance imaging has been modeled through both brain-network representations and temporal sequence learning. BrainNetCNN explicitly operates on functional-connectivity matrices through specialized edge-to-edge convolutions \citep{kawahara2017brainnetcnn}, while BrainGNN models brain regions and their interactions as a graph to learn disease-related network patterns \citep{li2021braingnn}. Brain Network Transformer further uses Transformer-based interactions to capture long-range dependencies among brain regions \citep{kan2022brainnetworktransformer}. In contrast to connectivity-based representations, sequence-based approaches retain the temporal dynamics of BOLD signals. BrainLM, for example, employs large-scale self-supervised pretraining on brain time series to learn transferable representations \citep{caro2023brainlm}. These methods substantially strengthen patient representation learning, but remain predominantly data driven.

External medical knowledge provides a complementary source of diagnostic information. Medical knowledge graphs and language models have been explored for knowledge retrieval, clinical question answering, and knowledge-grounded prediction \citep{chandak2023primekg,singhal2023clinical,thirunavukarasu2023llm}. Post hoc approaches such as Grad-CAM and SHAP can further identify influential imaging regions or clinical variables \citep{selvaraju2017gradcam,lundberg2017unified}. However, such knowledge is typically used as contextual information or retrospective explanation rather than as explicit structure for differential diagnosis. This motivates explicit patient-grounded knowledge structures for differential diagnosis.
\section{Method}
\label{sec:method}

\begin{figure}[!htbp]
    \centering
    \includegraphics[width=\textwidth,trim=8bp 10bp 4bp 14bp,clip]{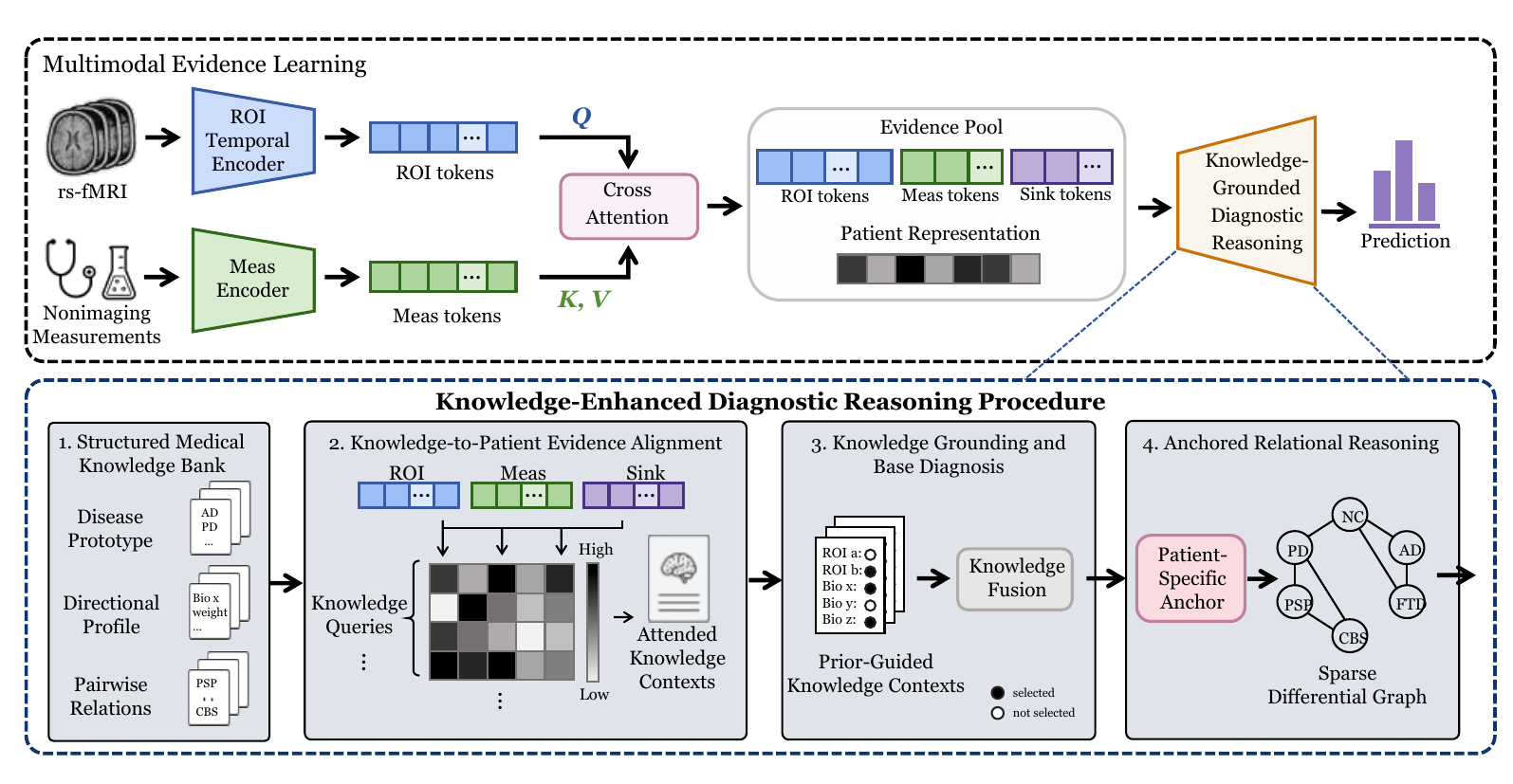}
    \caption{Overview of KnowSR for six-class neurodegenerative differential diagnosis. ROI-wise BOLD signals are encoded by a shared temporal encoder, while nonimaging measurements are projected into measurement tokens and integrated with ROI representations through cross-attention to obtain a multimodal patient representation and token-level evidence pool. A fixed structured medical knowledge bank containing disease prototypes, directional profiles, and pairwise relations is then aligned with the patient evidence to obtain patient-specific knowledge contexts. These contexts are fused to form a base diagnosis, which is further refined through a patient-specific diagnostic anchor and sparse differential graph for the final prediction.}
    \label{fig:knowsr_overview}
\end{figure}

In the following, we will first present the notations used in our method, and then describe the key modules of our knowledge-enhanced structural reasoning as shown in Fig.~\ref{fig:knowsr_overview}.
% KnowSR comprises three stages, as illustrated in Fig.~\ref{fig:knowsr_overview}. First, a temporal encoder converts ROI-wise BOLD sequences into regional tokens and conditions them on available nonimaging measurements. Second, a fixed structured knowledge bank is aligned with patient evidence to produce knowledge-grounded base logits. Third, disease-specific, directional, and pairwise evidence refine these logits through anchored relational projection. The following subsections detail the representation, knowledge bank, grounding, relational reasoning, and training objective.
%定义多模态输入输出，Assume，系统大致架构x y F

\subsection{Preliminary}
\label{sec:problem_formulation}

We formulate neurodegenerative differential diagnosis as a six-class classification problem. Let $R=132$, $M=51$, and $C=6$ denote the numbers of brain regions, nonimaging measurements, and diagnostic classes, respectively.

For scan $i$, the multimodal patient input is $\mathbf{x}_i=(\mathbf{x}_i^{R},\mathbf{x}_i^{M},\mathbf{o}_i)$, where $\mathbf{x}_i^{R}\in\mathbb{R}^{T_i\times R}$ denotes the ROI-wise BOLD sequence with $T_i$ time points, $\mathbf{x}_i^{M}\in\mathbb{R}^{M}$ contains the nonimaging measurements after feature-wise standardization using statistics estimated exclusively from the training partition, and $\mathbf{o}_i\in\{0,1\}^{M}$ indicates their availability. The availability mask excludes unobserved measurements from subsequent computation and is not encoded as an independent predictive feature.

The diagnostic label is $y_i\in\mathcal{Y}$, with $\mathcal{Y}=\{\mathrm{NC},\mathrm{AD},\mathrm{PD},\mathrm{FTD},\mathrm{PSP},\mathrm{CBS}\}$. Given $\mathbf{x}_i$ and the fixed structured medical knowledge bank $\mathcal{K}$, KnowSR produces scan-level logits $\mathbf{z}_i\in\mathbb{R}^{C}$ and probabilities $\mathbf{p}_i=\operatorname{softmax}(\mathbf{z}_i)$. For patients with repeated scans, patient-level probabilities are obtained by averaging $\mathbf{p}_i$ over their available scans.

%todo 不要太多符号，减少脚标,变量之间加逗号 ,尽量少用notation。从x到y到f，串起来。
%index尽量用常见的符号如ijk

\subsection{Multimodal Patient Evidence Encoding}
\label{sec:patient_encoding}

The patient branch maps the ROI-wise BOLD signals and nonimaging measurements into a common $d$-dimensional representation space. The regional time series follow the Schaefer-100 cortical and Tian S2 subcortical atlases. Let $\mathbf{x}_{i,j}^{R}\in\mathbb{R}^{T_i}$ denote the BOLD sequence of the $j$-th ROI in scan $i$. A shared temporal encoder applies three dilated one-dimensional convolutions with kernel sizes $(7,5,3)$ and dilations $(1,2,4)$, followed by learned temporal-position embeddings and attention pooling:
\begin{equation}
\mathbf{r}_{i,j}
=
\operatorname{LN}\!\left[
\mathbf{W}_{R}
\sum_{t=1}^{T_i}
\alpha_{i,j,t}
\left(
\mathcal{C}(\mathbf{x}_{i,j}^{R})_t
+
\mathbf{e}_{t}^{\mathrm{time}}
\right)
+
\mathbf{e}_{j}^{\mathrm{ROI}}
\right],
\label{eq:roi_encoding}
\end{equation}
where $\operatorname{LN}$ denotes layer normalization, $\mathcal{C}$ is the shared temporal convolutional encoder, $\mathbf{W}_{R}\in\mathbb{R}^{d\times h}$ is the learned projection from the $h$-dimensional temporal feature space to the common representation space, $\mathbf{e}_{t}^{\mathrm{time}}$ is the learned temporal-position embedding, $\alpha_{i,j,t}$ is the normalized attention weight over the valid time points of ROI $j$ in scan $i$, and $\mathbf{e}_{j}^{\mathrm{ROI}}$ is the learned ROI-identity embedding. The same encoder is applied independently to all ROIs without imposing anatomical adjacency.

Nonimaging measurements are mapped into the same $d$-dimensional space by a shared scalar encoder augmented with learned measurement-identity and type embeddings. For unavailable measurements, the scalar input is zeroed before projection, and the availability vector $\mathbf{o}_i$ prevents the corresponding tokens from participating as keys or values in subsequent attention. Let $\mathbf{R}_i$ and $\mathbf{M}_i$ denote the resulting ROI and measurement token sets, respectively. Cross-modal fusion uses multi-head attention (MHA), with the ROI tokens as queries and the available measurement tokens as keys and values:
\begin{equation}
\widetilde{\mathbf{R}}_i
=
\operatorname{MHA}
\left(
\mathbf{R}_i,
\mathbf{M}_i,
\mathbf{M}_i;
\mathbf{o}_i
\right),
\qquad
\mathbf{f}_i
=
\operatorname{MeanPool}
\left(
\widetilde{\mathbf{R}}_i
\right).
\label{eq:multimodal_fusion}
\end{equation}
When no nonimaging measurement is available, we set $\widetilde{\mathbf{R}}_i=\mathbf{R}_i$ and directly pool the original ROI tokens. For subsequent Card-to-patient grounding, the measurement-conditioned ROI tokens $\widetilde{\mathbf{R}}_i$, measurement tokens $\mathbf{M}_i$, and six learned sink tokens form the token-level patient evidence pool $\mathbf{H}_i$. The sink tokens provide null positions for unmatched Card queries, while unavailable measurement tokens remain masked.

\subsection{Structured Medical Knowledge Bank}
\label{sec:knowledge_bank}

In parallel with patient encoding, medical prior knowledge is organized into a fixed literature-grounded bank $\mathcal{K}$ containing 36 structured Cards: 23 disease-prototype Cards, 6 disease-direction Cards, and 7 pairwise-relation Cards. Each Card is represented as
$\mathbf{c}_k=[\mathbf{c}_{k,\mathrm{Diag}};
\mathbf{c}_{k,\mathrm{ROI}};
\mathbf{c}_{k,\mathrm{Meas}};
c_{k,\mathrm{Prio}}]$,
where $\mathbf{c}_{k,\mathrm{Diag}}\in\mathbb{R}^{6}$ identifies the associated diagnosis or diagnostic pair, $\mathbf{c}_{k,\mathrm{ROI}}\in\mathbb{R}^{132}$ specifies the predefined ROI targets, $\mathbf{c}_{k,\mathrm{Meas}}\in\mathbb{R}^{51}$ specifies the targeted nonimaging measurements and their encoded evidence attributes, and $c_{k,\mathrm{Prio}}$ represents the Card priority.

The Cards were constructed offline as a literature-grounded external knowledge bank through an LLM-assisted literature synthesis process. The LLM was used to summarize recurring anatomical, clinical, and biomarker patterns from publications covering the six diagnostic categories and to organize them into the predefined prototype, disease-direction, and pairwise schemas. A retrospective source audit linked the finalized bank to 38 supporting publications. The extracted anatomical terms were mapped to the fixed 132-ROI vocabulary, and clinical and biomarker findings were aligned with the predefined 51-measurement panel; unspecified entries were set to zero. For disease-direction and pairwise Cards, nonzero measurement entries encode the expected direction and a predefined ordinal strength, while prototype Cards use fixed priority weights. The finalized Cards were frozen before the reported model fitting and used unchanged across all folds, with no online retrieval or fold-specific adaptation.

The ROI, measurement, and priority components further determine a nonnegative target prior $\mathbf{a}_k$ over the patient-evidence tokens. Its nonzero entries indicate the predefined medical targets of Card $k$, while six zero entries are appended at the sink-token positions to match the patient evidence pool $\mathbf{H}_i$.

\subsection{Patient Grounding and Base Diagnosis}
\label{sec:knowledge_alignment}

The patient and knowledge branches are coupled through Card-to-patient grounding. Each structured Card is first mapped into the patient representation space and then queries the token-level patient evidence pool $\mathbf{H}_i$ through multi-head cross-attention, with its predefined target prior providing guidance toward medically relevant evidence: 
\begin{equation}
\mathbf{g}_{i,k}
=
\operatorname{CrossAttn}
\left(
\operatorname{Proj}(\mathbf{c}_k),
\mathbf{H}_i;
\mathbf{a}_k
\right),
\label{eq:knowledge_attention}
\end{equation}
where $\operatorname{Proj}(\cdot)$ maps Card $\mathbf{c}_k$ into the $d$-dimensional patient representation space, and $\mathbf{g}_{i,k}$ denotes the representation of Card $k$ grounded in the evidence of scan $i$. The projected Card serves as the query, while $\mathbf{H}_i$ provides the keys and values. The target prior $\mathbf{a}_k$ guides the alignment toward predefined medical targets, while unavailable measurements are excluded from attention.

The grounded Card representations are aggregated across the knowledge bank and combined with a prior-guided context derived directly from the patient evidence pool. Together with the pooled multimodal patient representation $\mathbf{f}_i$, they form the base diagnostic logits:
\begin{equation}
\mathbf{z}_{i}^{0}
=
\operatorname{MLP}\!\left[
\mathbf{f}_i
+
\beta
\left(
\operatorname{Agg}_{k}
\left(
\mathbf{g}_{i,k}
\right)
+
\operatorname{Agg}^{\mathrm{prior}}_{k}
\left(
\mathbf{H}_i;\mathbf{a}_k
\right)
\right)
\right].
\label{eq:base_diagnosis}
\end{equation}
Here, $\mathbf{f}_i$ provides the direct patient-derived representation,
$\operatorname{Agg}_{k}(\cdot)$ summarizes the patient-grounded Card
representations, and
$\operatorname{Agg}^{\mathrm{prior}}_{k}(\mathbf{H}_i;\mathbf{a}_k)$
summarizes the available patient evidence selected by the predefined Card
target priors. Both terms are aggregated over the three Card groups using
bounded group-wise gates, while $\beta$ controls their overall contribution
to the base diagnosis. The resulting $\mathbf{z}_{i}^{0}$ serves as the base
diagnosis for subsequent structural reasoning.

\subsection{Anchored Relational Reasoning}
\label{sec:relational_reasoning}

The grounded Cards provide two complementary class-wise corrections to the base diagnosis. Prototype-based evidence $\mathbf{e}_{i}^{\mathrm{Proto}}\in\mathbb{R}^{C}$ measures how strongly the patient-grounded representations match the characteristic prototype patterns of each diagnosis. Direction-based evidence $\mathbf{e}_{i}^{\mathrm{Dir}}\in\mathbb{R}^{C}$ evaluates whether the observed nonimaging measurements follow the expected disease-specific directions, with positive and negative values indicating supportive and contradictory evidence, respectively. The two evidence terms are adaptively weighted before refining the knowledge-grounded base diagnosis into a patient-specific diagnostic anchor:
\begin{equation}
\mathbf{z}_{i}^{A}
=
\mathbf{z}_{i}^{0}
+
\widetilde{\mathbf{e}}_{i}^{\mathrm{Proto}}
+
\widetilde{\mathbf{e}}_{i}^{\mathrm{Dir}},
\label{eq:disease_anchor}
\end{equation}
where $\widetilde{\mathbf{e}}_{i}^{\mathrm{Proto}}$ and
$\widetilde{\mathbf{e}}_{i}^{\mathrm{Dir}}$ denote the adaptively weighted
prototype-based and direction-based evidence, respectively, and
$\mathbf{z}_{i}^{A}$ denotes the patient-specific diagnostic anchor used for subsequent
relational reasoning.

Pairwise Cards further model relative evidence between clinically competing diagnoses. Rather than constructing an exhaustive all-pairs graph, the seven pairwise relations are organized according to clinically meaningful local differential structures. NC, AD, and FTD form a cognitive--dementia neighborhood, while PD, PSP, and CBS form a parkinsonian neighborhood; all pairwise edges within each neighborhood are retained to capture local diagnostic ambiguity. A single NC--PD edge connects the two neighborhoods, yielding a sparse but globally connected graph.

For each edge $e$, let $\boldsymbol{\delta}_e$ denote the signed measurement profile derived from the corresponding pairwise Card, encoding both the direction and strength of its measurement evidence. Comparing this profile with the standardized measurements observed in scan $i$ gives the pairwise margin
$m_{i,e}=\operatorname{PairScore}(\mathbf{x}_i^M,\mathbf{o}_i,\boldsymbol{\delta}_e)$, 
where $\operatorname{PairScore}(\cdot)$ computes an observed-strength-normalized signed aggregation over the available measurements specified by $\boldsymbol{\delta}_e$, followed by a learned relation-specific scale and bias. Unavailable measurements are excluded from the aggregation. The sign of $m_{i,e}$ indicates which endpoint of edge $e$ is favored, while its magnitude reflects the strength of the available evidence. Each active edge is further assigned a projection confidence $\rho_{i,e}$ that combines its learned relation gate, its relevance to the current base diagnosis, and the coverage of its observed evidence.

For auxiliary knowledge supervision, the pairwise margins are additionally mapped back to the shared six-class space:
\begin{equation}
\mathbf{e}_{i}^{\mathrm{Pair}}
=
\sum_{e=1}^{7}
\omega_{i,e}
m_{i,e}
\mathbf{b}_{e},
\qquad
\mathbf{z}_{i}^{E}
=
\mathbf{e}_{i}^{\mathrm{Proto}}
+
\mathbf{e}_{i}^{\mathrm{Dir}}
+
\mathbf{e}_{i}^{\mathrm{Pair}},
\label{eq:structured_evidence_prediction}
\end{equation}
where $\mathbf{b}_{e}$ encodes the orientation of edge $e$, and $\omega_{i,e}$ combines the corresponding Pair Card priority, its relevance to the current base diagnosis, and the coverage of its observed evidence. The resulting $\mathbf{z}_{i}^{E}$ represents the structured-evidence prediction used for auxiliary supervision in Section~\ref{sec:training_objective}.

The patient-specific diagnostic anchor and pairwise evidence are finally reconciled in the shared diagnostic space:
\begin{equation}
\mathbf{z}_i
=
\arg\min_{\mathbf{z}\in\mathbb{R}^{C}}
\left[
\eta
\left\|
\mathbf{z}
-
\mathbf{z}_{i}^{A}
\right\|_2^2
+
\sum_{e=1}^{7}
\rho_{i,e}
\left(
\mathbf{b}_{e}^{\top}\mathbf{z}
-
m_{i,e}
\right)^2
\right],
\label{eq:graph_projection}
\end{equation}
where $\mathbf{b}_{e}$ encodes the orientation of edge $e$, such that $\mathbf{b}_{e}^{\top}\mathbf{z}$ represents the diagnostic logit difference between its two endpoints. The first term preserves the patient-specific diagnostic anchor, while the second adjusts pairwise diagnostic differences toward the evidence-supported margins. The coefficient $\eta$ controls the balance between direct disease-level evidence and pairwise relational refinement. The resulting $\mathbf{z}_i$ gives the final scan-level logits, with class probabilities obtained by $\operatorname{softmax}(\mathbf{z}_i)$.

\subsection{Training Objective}
\label{sec:training_objective}

All trainable components are jointly optimized by
\begin{equation}
\mathcal{L}
=
\mathcal{L}_{\mathrm{cls}}
+
\lambda_{\mathrm{diag}}\mathcal{L}_{\mathrm{diag}}
+
\lambda_{\mathrm{know}}\mathcal{L}_{\mathrm{know}},
\label{eq:training_objective}
\end{equation}
where $\mathcal{L}_{\mathrm{cls}}$ is the primary classification loss,
$\mathcal{L}_{\mathrm{diag}}$ provides auxiliary scan- and patient-level supervision,
and $\mathcal{L}_{\mathrm{know}}$ supervises intermediate knowledge-guided predictions.

The primary term is the class-weighted, label-smoothed cross-entropy
$\mathcal{L}_{\mathrm{cls}}=\operatorname{CE}_{w,\epsilon}(\mathbf{z},y)$,
with class weights computed from patient counts in the training partition.
Each scan is additionally weighted by the inverse number of scans from the corresponding patient.

The diagnostic auxiliary loss combines class-balanced scan-level and patient-level supervision:
\begin{equation}
\mathcal{L}_{\mathrm{diag}}
=
-\frac{1}{2|\mathcal{B}|}
\sum_{i\in\mathcal{B}}
\tilde{w}_{y_i}\log p_{i,y_i}
+
\frac{
\sum_{u\in\mathcal{U}_{\mathcal{B}}}
w_{y_u}\ell_{\epsilon}(\bar{\mathbf{p}}_u,y_u)
}{
2\sum_{u\in\mathcal{U}_{\mathcal{B}}}w_{y_u}
},
\label{eq:diagnostic_auxiliary}
\end{equation}
where $\mathcal{B}$ is a scan mini-batch,
$\mathcal{U}_{\mathcal{B}}$ denotes the patients represented in the batch,
$\mathcal{I}_u$ contains the scans of patient $u$ within $\mathcal{B}$,
$y_u$ is the diagnosis of patient $u$, and
$\bar{\mathbf{p}}_u=|\mathcal{I}_u|^{-1}
\sum_{i\in\mathcal{I}_u}\mathbf{p}_i$
is the corresponding mean prediction.
Here, $w_y$ is the six-class weight derived from training-patient counts,
while $\tilde{w}_y$ averages the normalized inverse-frequency weights
across the predefined diagnostic label pairs containing class $y$.
The term $\ell_{\epsilon}$ denotes label-smoothed negative log-likelihood
obtained by mixing the one-hot target with the uniform class distribution,
with $\epsilon=0.02$.

The knowledge term is
$\mathcal{L}_{\mathrm{know}}
=\frac{1}{2}
[\operatorname{CE}_{w,\epsilon}(\mathbf{z}^{0},y)
+\operatorname{CE}_{w,\epsilon}(\mathbf{z}^{E},y)]$,
where $\mathbf{z}^{0}$ and $\mathbf{z}^{E}$ denote the base and structured-evidence predictions, respectively.
Both auxiliary losses are used only during training, while inference uses the final refined prediction $\mathbf{z}$.
\section{EXPERIMENTS}
\subsection{Study Cohorts and Data Preparation}
\label{sec:study_cohorts}

\begin{figure}[!htbp]
    \centering
    \includegraphics[width=0.98\textwidth]{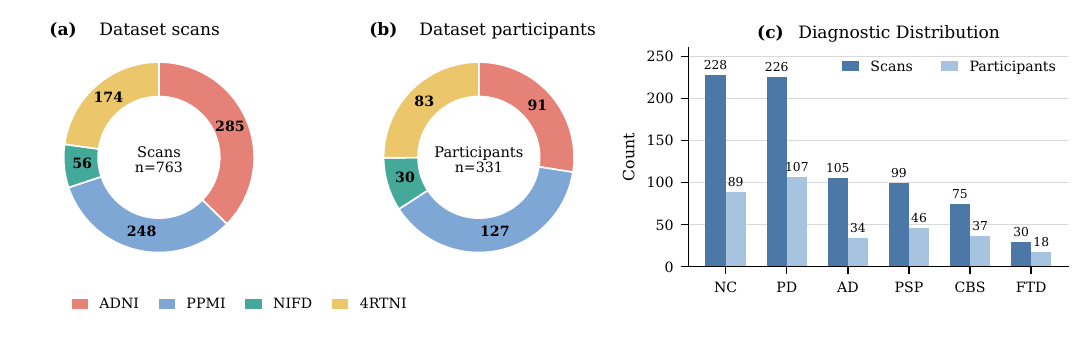}
    \caption{Study-cohort composition. (a) Scan-level and (b) participant-level contributions of the four databases; (c) scan- and participant-level distributions across the six diagnostic classes.}
    \label{fig:data_composition}
\end{figure}

We used four controlled-access research databases: the Alzheimer's Disease Neuroimaging Initiative (ADNI), Parkinson's Progression Markers Initiative (PPMI), Frontotemporal Lobar Degeneration Neuroimaging Initiative (NIFD), and 4 Repeat Tauopathy Neuroimaging Initiative (4RTNI)~\citep{adni_database,ppmi_database,nifd_database,rtni_database}. The final cohort comprised 331 participants and 763 resting-state functional magnetic resonance imaging (rs-fMRI) scans, with diagnostic coverage of NC/AD in ADNI, NC/PD in PPMI, NC/FTD in NIFD, and PSP/CBS in 4RTNI. Participants with repeated scans were treated as single individuals with multiple observations. Detailed database and diagnostic compositions are shown in Fig.~\ref{fig:data_composition}.

All rs-fMRI data were processed using a common pipeline. DICOM images were converted to NIfTI using dcm2niix. Slice-timing correction was applied with FSL \texttt{slicetimer} when timing information was available, followed by motion realignment with MCFLIRT. Mean functional images were registered to brain-extracted T1-weighted images using six-degree-of-freedom FLIRT. T1-weighted images were then normalized to the 2-mm MNI152 template using 12-degree-of-freedom affine registration followed by nonlinear FNIRT, and the resulting transformations were applied to the BOLD series.
Regional time series were extracted in MNI space using a 132-region atlas comprising 32 Tian S2 subcortical regions and 100 seven-network Schaefer cortical parcels, followed by linear detrending and standardization. The first and last five frames were removed, and scans with fewer than 100 remaining time points were excluded. Variable-length sequences were right-padded within each mini-batch, with padded positions masked during temporal attention and pooling. All retained sequences were within the 512-frame limit of the temporal-position embedding and were input directly without conversion to static functional-connectivity matrices.

Each scan was paired with 51 nonimaging measurements covering demographic, molecular, cognitive, behavioral, functional, and motor information, together with feature-level availability indicators. All data-dependent normalization statistics were estimated exclusively from the training partition of each fold and applied unchanged to the corresponding validation and test partitions. Unavailable measurements were assigned a neutral numeric placeholder during data loading and excluded from cross-modal attention and evidence aggregation using their availability indicators. The indicators themselves were never projected into learned representations or concatenated with model inputs, ensuring that missing measurements contributed no patient evidence and were not interpreted as observed zeros.

\subsection{Evaluation Protocol}
\label{sec:evaluation_protocol}

All experiments used patient-disjoint five-fold evaluation, with all scans from the same participant assigned to the same fold. Each fold contained training, validation, and test subsets, and every participant appeared in the held-out test set exactly once. Identical partitions were used for all compared methods. The five test folds contained 65--67 participants and 141--162 scans. Hyperparameter tuning and checkpoint selection used only the validation subset, with no test-set involvement.

For patient-level prediction, unrestricted six-class scan probabilities were averaged across all available scans from the same participant, and the class with the highest mean probability was selected. Performance was evaluated using balanced accuracy (BA), macro-F1, and one-versus-rest macro-AUC, which respectively average class-wise recall, F1 score, and AUC across the six diagnostic classes.

Each metric was computed independently on the held-out test subset of each fold and reported as the five-fold mean and sample standard deviation. Pooled out-of-fold analyses were performed only after concatenating each participant's single held-out prediction and were reported separately.

We additionally evaluated within-database discrimination for NC versus AD in ADNI, NC versus PD in PPMI, NC versus FTD in NIFD, and PSP versus CBS in 4RTNI. These analyses used only the corresponding two entries of the unrestricted six-class predictions; database identity was never used to constrain model outputs during inference.

\subsection{Compared Methods}
\label{sec:compared_methods}

We compared KnowSR with 19 baselines spanning four families: traditional machine-learning methods (RBF-SVM~\citep{cortes1995support}, logistic regression~\citep{cox1958regression}, random forest~\citep{breiman2001random}, and XGBoost~\citep{chen2016xgboost}); generic sequence and multimodal models (Contrasformer~\citep{xu2024contrasformer}, BiLSTM~\citep{schuster1997bidirectional}, and MulT~\citep{tsai2019multimodal}); brain-network models (BNT~\citep{kan2022brainnetworktransformer}, STAGIN~\citep{kim2021stagin}, BrainNetCNN~\citep{kawahara2017brainnetcnn}, and BrainGNN~\citep{li2021braingnn}); and medical multimodal models (DHGFormer~\citep{xue2025dhgformer}, CAGT~\citep{pei2025cagt}, DrFuse~\citep{yao2024drfuse}, MedFuse~\citep{hayat2022medfuse}, ContrastPool~\citep{xu2024contrastpool}, GDL~\citep{zhang2025gdl}, BrainLM~\citep{caro2023brainlm}, and NMED24~\citep{xue2024ai}). Public implementations were adapted to the common 132-ROI representation, 51-measurement panel, and six-class prediction space. BrainLM retained its released pretrained encoder and was fine-tuned separately within each training fold; all other neural baselines were trained from the study folds without test-set access.

For classical models, the upper triangle of the 132-ROI static functional-connectivity matrix (8,646 edges) was concatenated with the 51 measurements, with missing values deterministically imputed using training-fold statistics. Neural baselines retained their native temporal, graph, or multimodal encoders and used structurally masked measurement inputs without exposing availability indicators as learned features. All methods used identical patient-disjoint folds, seed 42, and the patient-level aggregation defined in Section~\ref{sec:evaluation_protocol}. Hyperparameters and, where applicable, neural checkpoints were selected exclusively by validation balanced accuracy, with no test-set involvement.

\subsection{Implementation Details}
\label{sec:implementation_details}

The temporal encoder used three one-dimensional convolutional blocks with 64 channels, kernel sizes $(7,5,3)$, and dilations $(1,2,4)$. KnowSR used a token dimension of $d=256$, four attention heads, and dropout 0.1. The Card-prior coefficient was $\alpha=0.7$, the Card-group gate bound $\gamma_{\max}=0.3$, the knowledge-fusion coefficient $\beta=0.3$, and the disease-wise evidence-gate bound $g_{\max}=0.3$. Pairwise evidence was clipped at $\kappa=5$, relation-specific scale and bias were bounded by $a_{\max}=15$, and the anchor coefficient was set to $\eta=1.0$.

All trainable parameters were randomly initialized and jointly optimized using AdamW with a learning rate of $3\times10^{-5}$, weight decay $10^{-4}$, label smoothing 0.02, and gradient-norm clipping at 5. Relation-specific scale, bias, and gate parameters used a $15\times$ learning-rate multiplier. Patient-grouped mini-batches had a nominal size of 8, and the loss weights were $\lambda_{\mathrm{diag}}=0.4$ and $\lambda_{\mathrm{know}}=0.6$. The four auxiliary diagnostic subspaces were NC--AD, NC--PD, NC--FTD, and PSP--CBS, with each sample contributing to every subspace containing its ground-truth class.

Training ran for at most 80 epochs with early stopping after 20 epochs without improvement in validation balanced accuracy. One checkpoint per fold was retained according to the highest validation balanced accuracy. Random seed 42 was used throughout. No pretrained weights, checkpoint averaging, or test-time adaptation were used. The five folds were trained concurrently on NVIDIA GeForce RTX 4090 GPUs, with one fold assigned to each GPU.

\section{Results and Discussion}
\subsection{Overall Differential Diagnosis Performance}
\label{sec:overall_results}

KnowSR achieved the best patient-level six-class performance among all 19 baselines, reaching $91.94\%\pm4.33\%$ balanced accuracy, $90.75\%\pm4.93\%$ macro-F1, and $99.02\%\pm0.72\%$ OVR macro-AUC (Table~\ref{tab:overall_patient}). These results exceeded the strongest baseline by 6.27, 6.08, and 1.13 percentage points, respectively. Paired analysis over the 331 out-of-fold patients with 10,000 stratified bootstrap resamples further showed significant improvements over both NMED24 and BrainLM, with all 95\% confidence intervals excluding zero. Compared with NMED24, KnowSR improved balanced accuracy, macro-F1, and OVR macro-AUC by 6.16 points (2.12--10.29), 6.18 points (2.09--10.35), and 1.83 points (1.01--2.71), respectively; the corresponding gains over BrainLM were 6.44 (2.21--10.66), 6.30 (2.25--10.44), and 1.19 points (0.50--1.97).

\begin{table}[!htbp]
\centering
\caption{Overall patient-level six-class performance.}
\label{tab:overall_patient}

\small
\setlength{\tabcolsep}{12pt}
\renewcommand{\arraystretch}{1.12}

\begin{tabular}{@{}lccc@{}}
\toprule
Method & BA & Macro-F1 & OVR-AUC \\
\midrule

\multicolumn{4}{@{}l}{\textit{Traditional ML}} \\
\addlinespace[2pt]

RBF-SVM
& $44.10 \pm 7.76$
& $43.59 \pm 8.99$
& $85.26 \pm 2.12$ \\

Logistic Regression
& $50.93 \pm 3.80$
& $50.99 \pm 4.99$
& $81.71 \pm 2.23$ \\

Random Forest
& $52.19 \pm 3.37$
& $54.97 \pm 2.81$
& $89.73 \pm 3.14$ \\

XGBoost
& $64.00 \pm 8.64$
& $66.23 \pm 8.39$
& $92.40 \pm 2.27$ \\

\addlinespace[5pt]
\multicolumn{4}{@{}l}{\textit{Generic DL}} \\
\addlinespace[2pt]

Contrasformer
& $59.91 \pm 8.39$
& $59.21 \pm 8.60$
& $92.70 \pm 3.15$ \\

BiLSTM
& $63.45 \pm 5.76$
& $63.89 \pm 5.99$
& $94.54 \pm 1.78$ \\

MulT
& $65.52 \pm 8.86$
& $63.38 \pm 9.69$
& $93.65 \pm 2.76$ \\

\addlinespace[5pt]
\multicolumn{4}{@{}l}{\textit{Brain network/graph}} \\
\addlinespace[2pt]

BNT
& $63.49 \pm 6.95$
& $61.34 \pm 5.31$
& $93.66 \pm 1.73$ \\

STAGIN
& $64.82 \pm 7.29$
& $64.54 \pm 7.67$
& $94.44 \pm 2.08$ \\

BrainNetCNN
& $68.55 \pm 9.47$
& $66.03 \pm 9.64$
& $95.90 \pm 1.89$ \\

BrainGNN
& $79.16 \pm 1.30$
& $76.90 \pm 2.97$
& $97.09 \pm 0.92$ \\

\addlinespace[5pt]
\multicolumn{4}{@{}l}{\textit{Medical multimodal}} \\
\addlinespace[2pt]

DHGFormer
& $55.35 \pm 4.67$
& $53.92 \pm 4.54$
& $90.67 \pm 1.88$ \\

CAGT
& $55.83 \pm 6.46$
& $51.85 \pm 3.79$
& $89.30 \pm 1.39$ \\

DrFuse
& $72.10 \pm 7.33$
& $71.26 \pm 6.31$
& $94.78 \pm 2.06$ \\

MedFuse
& $75.37 \pm 7.23$
& $72.15 \pm 9.32$
& $96.54 \pm 0.47$ \\

ContrastPool
& $78.04 \pm 2.76$
& $76.32 \pm 2.74$
& $97.26 \pm 0.75$ \\

GDL
& $78.98 \pm 3.16$
& $77.92 \pm 3.71$
& $97.37 \pm 1.06$ \\

BrainLM
& $85.67 \pm 2.33$
& $84.61 \pm 2.33$
& $97.89 \pm 0.67$ \\

NMED24
& $85.51 \pm 2.96$
& $84.67 \pm 4.66$
& $97.23 \pm 1.29$ \\

\midrule

\textbf{KnowSR (Ours)}
& $\mathbf{91.94}\pm\mathbf{4.33}$
& $\mathbf{90.75}\pm\mathbf{4.93}$
& $\mathbf{99.02}\pm\mathbf{0.72}$ \\

\bottomrule
\end{tabular}

\end{table}

\newgeometry{top=36bp,bottom=36bp,left=54bp,right=54bp,footskip=18bp}
\begin{landscape}
\begin{table}[p]
\caption{Patient-level performance within individual datasets.}
\label{tab:patient_dataset_results}
\centering
\footnotesize
\setlength{\tabcolsep}{2.2pt}
\renewcommand{\arraystretch}{1.08}
\resizebox{\linewidth}{!}{%
\begin{tabular}{lcccccccccccc}
\toprule
\multirow{2}{*}{Method} & \multicolumn{3}{c}{ADNI} & \multicolumn{3}{c}{PPMI} & \multicolumn{3}{c}{NIFD} & \multicolumn{3}{c}{4RTNI} \\
\cmidrule(lr){2-4}\cmidrule(lr){5-7}\cmidrule(lr){8-10}\cmidrule(lr){11-13}
& BA & Macro-F1 & AUC & BA & Macro-F1 & AUC & BA & Macro-F1 & AUC & BA & Macro-F1 & AUC \\
\midrule
\multicolumn{13}{l}{\textit{Traditional machine learning}} \\
RBF-SVM & $61.09{\pm}12.07$ & $60.94{\pm}13.78$ & $72.52{\pm}13.15$ & $49.21{\pm}6.45$ & $47.56{\pm}8.23$ & $61.86{\pm}12.37$ & $71.67{\pm}19.85$ & $64.43{\pm}26.57$ & $69.44{\pm}22.57$ & $52.53{\pm}7.74$ & $51.27{\pm}6.30$ & $49.23{\pm}9.64$ \\
Logistic Regression & $64.75{\pm}5.73$ & $64.38{\pm}5.24$ & $65.88{\pm}10.23$ & $47.64{\pm}2.38$ & $44.51{\pm}1.29$ & $56.65{\pm}16.10$ & $83.33{\pm}10.21$ & $82.03{\pm}12.03$ & $86.11{\pm}14.96$ & $55.53{\pm}16.55$ & $53.89{\pm}15.24$ & $56.74{\pm}16.72$ \\
Random Forest & $92.02{\pm}3.76$ & $92.72{\pm}2.92$ & $99.22{\pm}0.62$ & $44.83{\pm}2.65$ & $43.00{\pm}1.54$ & $53.14{\pm}14.66$ & $78.33{\pm}15.42$ & $72.95{\pm}19.39$ & $85.83{\pm}19.45$ & $55.83{\pm}11.35$ & $53.28{\pm}11.77$ & $58.09{\pm}21.75$ \\
XGBoost & $97.74{\pm}3.27$ & $97.63{\pm}3.25$ & $98.10{\pm}3.63$ & $52.21{\pm}10.21$ & $49.70{\pm}11.51$ & $70.84{\pm}5.52$ & $88.33{\pm}16.24$ & $86.67{\pm}18.26$ & $97.78{\pm}4.97$ & $69.92{\pm}14.23$ & $69.32{\pm}15.20$ & $71.07{\pm}16.76$ \\
\addlinespace
\multicolumn{13}{l}{\textit{Generic deep learning}} \\
Contrasformer & $86.11{\pm}1.92$ & $85.86{\pm}2.74$ & $95.85{\pm}4.54$ & $64.05{\pm}13.97$ & $65.52{\pm}18.77$ & $73.48{\pm}28.43$ & $62.50{\pm}21.25$ & $57.63{\pm}24.71$ & $70.56{\pm}35.32$ & $51.14{\pm}7.84$ & $49.26{\pm}8.56$ & $52.78{\pm}7.67$ \\
BiLSTM & $87.19{\pm}4.27$ & $87.92{\pm}4.28$ & $95.59{\pm}6.34$ & $60.60{\pm}10.72$ & $60.35{\pm}15.09$ & $75.78{\pm}9.29$ & $81.67{\pm}14.01$ & $78.41{\pm}18.68$ & $95.56{\pm}9.94$ & $48.39{\pm}12.58$ & $47.21{\pm}13.12$ & $55.92{\pm}11.00$ \\
MulT & $89.42{\pm}7.67$ & $88.60{\pm}8.72$ & $95.51{\pm}5.34$ & $69.29{\pm}15.14$ & $66.05{\pm}16.80$ & $80.60{\pm}14.33$ & $80.83{\pm}21.97$ & $75.74{\pm}28.29$ & $93.33{\pm}14.91$ & $56.75{\pm}10.35$ & $55.26{\pm}10.04$ & $62.30{\pm}9.15$ \\
\addlinespace
\multicolumn{13}{l}{\textit{Brain network and graph learning}} \\
BNT & $86.71{\pm}4.97$ & $86.09{\pm}6.38$ & $92.94{\pm}3.85$ & $59.05{\pm}17.15$ & $58.62{\pm}20.81$ & $75.12{\pm}18.63$ & $59.17{\pm}14.55$ & $52.00{\pm}17.77$ & $81.11{\pm}22.08$ & $55.28{\pm}6.99$ & $54.29{\pm}6.94$ & $57.53{\pm}6.78$ \\
STAGIN & $82.94{\pm}8.25$ & $83.15{\pm}8.32$ & $89.38{\pm}4.84$ & $61.19{\pm}6.91$ & $60.68{\pm}8.56$ & $77.19{\pm}11.75$ & $66.67{\pm}21.85$ & $65.08{\pm}22.77$ & $81.11{\pm}22.08$ & $56.17{\pm}13.37$ & $55.69{\pm}13.02$ & $56.48{\pm}18.83$ \\
BrainNetCNN & $88.46{\pm}6.45$ & $88.33{\pm}6.74$ & $96.15{\pm}5.27$ & $59.64{\pm}10.38$ & $59.09{\pm}15.08$ & $79.09{\pm}12.47$ & $83.33{\pm}5.10$ & $81.84{\pm}2.27$ & $93.33{\pm}14.91$ & $50.19{\pm}8.73$ & $46.70{\pm}6.27$ & $58.44{\pm}10.31$ \\
BrainGNN & $91.71{\pm}1.97$ & $91.75{\pm}3.14$ & $96.89{\pm}4.27$ & $73.57{\pm}11.45$ & $76.00{\pm}10.41$ & $89.18{\pm}7.43$ & $72.50{\pm}14.91$ & $69.20{\pm}18.33$ & $92.50{\pm}16.77$ & $60.14{\pm}6.28$ & $58.76{\pm}6.47$ & $63.30{\pm}12.24$ \\
\addlinespace
\multicolumn{13}{l}{\textit{Medical multimodal and adapted models}} \\
DHGFormer & $84.77{\pm}7.74$ & $84.52{\pm}7.10$ & $91.25{\pm}4.97$ & $50.00{\pm}0.00$ & $45.72{\pm}0.10$ & $65.38{\pm}19.53$ & $60.83{\pm}10.87$ & $57.77{\pm}17.20$ & $64.72{\pm}28.49$ & $48.81{\pm}7.81$ & $46.93{\pm}8.81$ & $49.38{\pm}7.92$ \\
CAGT & $77.88{\pm}8.55$ & $75.57{\pm}9.81$ & $83.29{\pm}6.28$ & $47.64{\pm}2.92$ & $44.49{\pm}1.59$ & $47.15{\pm}17.28$ & $61.67{\pm}6.85$ & $55.21{\pm}10.43$ & $83.61{\pm}12.97$ & $50.19{\pm}9.93$ & $47.30{\pm}8.90$ & $50.73{\pm}9.80$ \\
DrFuse & $92.31{\pm}4.97$ & $92.79{\pm}5.08$ & $97.17{\pm}2.49$ & $59.05{\pm}8.84$ & $59.04{\pm}12.22$ & $77.34{\pm}13.93$ & $75.00{\pm}25.00$ & $70.22{\pm}32.03$ & $90.83{\pm}14.55$ & $46.67{\pm}6.35$ & $45.73{\pm}6.87$ & $49.17{\pm}11.60$ \\
MedFuse & $91.16{\pm}3.28$ & $90.66{\pm}3.23$ & $97.12{\pm}4.34$ & $57.02{\pm}6.49$ & $57.42{\pm}10.82$ & $83.21{\pm}13.40$ & $80.00{\pm}27.39$ & $74.67{\pm}34.77$ & $97.78{\pm}4.97$ & $52.31{\pm}12.05$ & $50.98{\pm}12.61$ & $53.92{\pm}9.20$ \\
ContrastPool & $87.71{\pm}4.24$ & $87.23{\pm}4.47$ & $95.36{\pm}5.06$ & $66.09{\pm}6.26$ & $69.41{\pm}7.14$ & $84.98{\pm}14.59$ & $83.33{\pm}15.59$ & $83.61{\pm}16.21$ & $97.50{\pm}5.59$ & $57.69{\pm}6.18$ & $56.31{\pm}5.64$ & $69.24{\pm}7.94$ \\
GDL & $89.52{\pm}3.22$ & $90.33{\pm}3.37$ & $96.11{\pm}5.20$ & $62.50{\pm}0.00$ & $66.72{\pm}0.08$ & $87.52{\pm}11.92$ & $88.33{\pm}11.18$ & $88.70{\pm}10.52$ & $\mathbf{100.00{\pm}0.00}$ & $58.67{\pm}6.43$ & $57.71{\pm}5.59$ & $64.34{\pm}12.62$ \\
BrainLM & $97.74{\pm}3.27$ & $97.63{\pm}3.25$ & $98.10{\pm}3.63$ & $81.55{\pm}13.89$ & $83.61{\pm}11.90$ & $91.73{\pm}9.16$ & $\mathbf{96.67{\pm}7.45}$ & $\mathbf{96.57{\pm}7.67}$ & $\mathbf{100.00{\pm}0.00}$ & $66.08{\pm}3.34$ & $64.69{\pm}2.87$ & $72.30{\pm}7.43$ \\
NMED24 & $96.31{\pm}6.19$ & $96.30{\pm}5.57$ & $98.91{\pm}1.81$ & $70.64{\pm}14.64$ & $70.50{\pm}15.44$ & $77.75{\pm}24.96$ & $90.00{\pm}22.36$ & $86.67{\pm}29.81$ & $\mathbf{100.00{\pm}0.00}$ & $75.08{\pm}6.21$ & $74.65{\pm}6.57$ & $78.48{\pm}10.97$ \\
\midrule
\textbf{KnowSR (Ours)} & $\mathbf{99.17{\pm}1.86}$ & $\mathbf{98.80{\pm}2.69}$ & $\mathbf{99.44{\pm}1.24}$ & $\mathbf{91.57{\pm}6.78}$ & $\mathbf{92.37{\pm}5.30}$ & $\mathbf{99.08{\pm}1.48}$ & $94.17{\pm}8.12$ & $93.14{\pm}9.39$ & $95.00{\pm}11.18$ & $\mathbf{83.72{\pm}8.06}$ & $\mathbf{82.67{\pm}7.95}$ & $\mathbf{86.61{\pm}10.94}$ \\
\bottomrule
\end{tabular}%
}
\begin{minipage}{\linewidth}
\end{minipage}
\end{table}
\end{landscape}
\restoregeometry

\subsection{Within-Database Differential Diagnosis}
\label{sec:database_results}

Within-database evaluation assessed whether KnowSR remained effective when discrimination was restricted to diagnostic groups acquired within the same database, thereby reducing potential confounding from between-database acquisition differences. As shown in Table~\ref{tab:patient_dataset_results}, KnowSR achieved the highest balanced accuracy in three of the four comparisons: $99.17\%$ for NC versus AD in ADNI, $91.57\%$ for NC versus PD in PPMI, and $83.72\%$ for PSP versus CBS in 4RTNI. In NIFD, KnowSR achieved $94.17\%$ balanced accuracy and $93.14\%$ macro-F1 for NC versus FTD, second to BrainLM at $96.67\%$ and $96.57\%$, respectively. For PSP versus CBS, KnowSR further exceeded NMED24 by 8.64 percentage points in balanced accuracy, 8.02 points in macro-F1, and 8.13 points in AUC. Together, these results indicate that KnowSR maintains strong diagnostic discrimination when comparisons are restricted within individual databases, reducing the influence of between-database acquisition differences.

To test whether database identity alone could account for KnowSR's performance, we conducted a patient-level permutation test as a negative control. Within each test fold and database, complete six-class prediction vectors were randomly reassigned among patients while preserving database membership, class composition, fold structure, and the unrestricted six-class decision rule. Thus, database-level prediction structure was retained, whereas the correspondence between predictions and individual patients was deliberately removed. Overall balanced accuracy and macro-F1 decreased from the observed $91.94\%$ and $90.75\%$ to $54.74{\pm}2.80\%$ and $53.89{\pm}2.77\%$, respectively, and none of the 1,000 permutations reproduced the observed performance ($p_{\mathrm{emp}}=9.99\times10^{-4}$). Within-database AUCs similarly approached chance level, reaching $50.07{\pm}6.55\%$ in ADNI, $50.12{\pm}7.05\%$ in PPMI, $48.64{\pm}11.57\%$ in NIFD, and $50.30{\pm}6.34\%$ in 4RTNI. Because performance markedly decreased after patient-level correspondence was disrupted while database membership and fold structure were preserved, these results indicate that database identity alone cannot account for KnowSR's diagnostic performance.

\subsection{Ablation and Evidence Analyses}
\label{sec:ablation}

\subsubsection{Component Ablations}

We examined whether KnowSR's gains arose from multimodal input alone or depended on cross-modal interaction, clinically meaningful knowledge, and anchored relational reasoning. Table~\ref{tab:patient_ablation} shows consistent contributions from these components.

\begin{table}[!htbp]
\caption{Patient-level component ablations and knowledge controls.}
\label{tab:patient_ablation}
\centering
\footnotesize
\setlength{\tabcolsep}{3.2pt}
\renewcommand{\arraystretch}{1.08}
\resizebox{\columnwidth}{!}{%
\begin{tabular}{lccc}
\toprule
Method & BA & Macro-F1 & OVR-AUC \\
\midrule
BOLD only & $51.97{\pm}4.61$ & $41.37{\pm}3.86$ & $89.11{\pm}1.16$ \\
Bio only & $76.39{\pm}6.75$ & $77.16{\pm}7.76$ & $92.92{\pm}2.53$ \\
Naive multimodal fusion & $75.69{\pm}2.32$ & $73.07{\pm}4.75$ & $96.78{\pm}1.11$ \\
BOLD + bio (cross-attention) & $85.70{\pm}5.24$ & $84.56{\pm}5.46$ & $98.25{\pm}0.77$ \\
Incorrect-Knowledge KnowSR & $87.50{\pm}5.16$ & $86.62{\pm}4.80$ & $98.41{\pm}0.70$ \\
KnowSR anchor-only & $86.60{\pm}5.48$ & $86.20{\pm}4.78$ & $98.24{\pm}0.76$ \\
\textbf{KnowSR (Ours)} & $\mathbf{91.94{\pm}4.33}$ & $\mathbf{90.75{\pm}4.93}$ & $\mathbf{99.02{\pm}0.72}$ \\
\bottomrule
\end{tabular}%
}
\end{table}

Cross-attention improved over naive fusion by 10.01 points in balanced accuracy, 11.49 in macro-F1, and 1.47 in OVR macro-AUC, showing that multimodal input alone was insufficient. Replacing the correct Card semantics with incorrect mappings, while preserving the architecture and number of knowledge units, reduced balanced accuracy and macro-F1 by 4.44 and 4.13 points, respectively. This control separates the effect of additional model structure from that of semantically aligned clinical knowledge and supports the value of disease-relevant evidence semantics.

Removing relational projection in Eq.~\eqref{eq:graph_projection} and predicting directly from the patient-specific diagnostic anchor $\mathbf{z}_i^A$ reduced balanced accuracy, macro-F1, and OVR macro-AUC from $91.94\%$, $90.75\%$, and $99.02\%$ to $86.60\%$, $86.20\%$, and $98.24\%$, respectively. Across 331 out-of-fold patients, 10,000 stratified bootstrap resamples estimated corresponding gains of 5.40 points in balanced accuracy (95\% CI, 2.18--9.04), 4.35 in macro-F1 (1.25--7.90), and 0.92 in OVR macro-AUC (0.40--1.51), with all intervals excluding zero. These results support the value of reconciling patient-specific pairwise evidence around patient-specific diagnostic anchors \dsyrev{for differential diagnosis.}

\subsubsection{Relation-Specific Clinical Refinement}

\begin{figure}[!htbp]
\centering
\includegraphics[width=0.99\textwidth,trim=5bp 11bp 0bp 3bp,clip]
{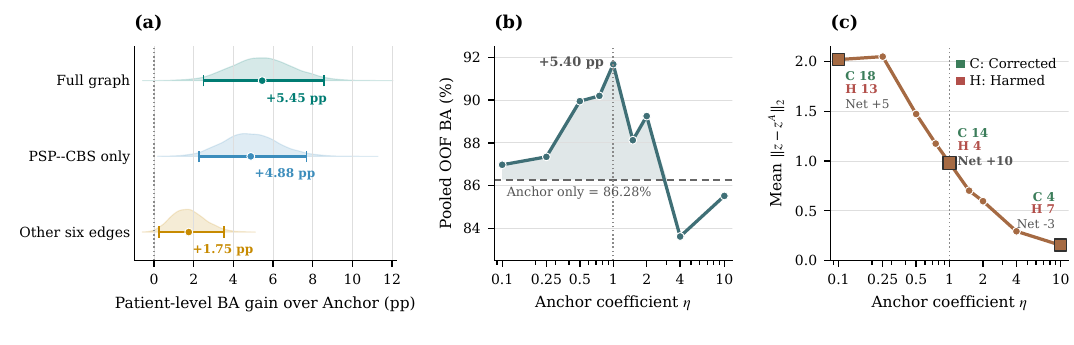}
\caption{Relational refinement and anchoring balance. (a) Relation-specific patient-level balanced-accuracy gain over the common pre-projection Anchor for the full seven-edge graph, the PSP--CBS edge alone, and the remaining six edges; points denote observed gains, half-violins show bootstrap distributions, and horizontal bars indicate 95\% confidence intervals. (b) Pooled out-of-fold patient-level balanced accuracy across nine anchoring coefficients $\eta$; the dashed line denotes the Anchor-only reference ($86.28\%$), the shaded region marks improvement above it, and the dotted line identifies the main setting ($\eta=1$). (c) Patient-balanced mean deviation from the anchor. Squares highlight $\eta=0.1$, $1$, and $10$; C and H denote corrected and harmed predictions relative to Anchor only, and Net denotes their difference.}
\label{fig:relational_refinement}

\end{figure}

To assess whether relational refinement was broadly distributed or driven by a clinically confusable pair, we performed inference-time edge ablations on the frozen full-model checkpoints, masking relation edges while keeping the same pre-projection anchor.
As shown in Fig.~\ref{fig:relational_refinement}(a), the full seven-edge graph improved balanced accuracy over the anchor by 5.45 points (95\% CI, 2.49--8.56). Retaining only the PSP--CBS edge yielded a 4.88-point gain (95\% CI, 2.25--7.69), recovering 89.5\% of the full improvement, whereas the remaining six edges contributed a smaller but positive gain of 1.75 points (95\% CI, 0.26--3.53). These results indicate that relational refinement is concentrated in the clinically challenging PSP--CBS differential while remaining complementary across the broader graph.

\subsubsection{Anchoring--Refinement Balance}

We varied $\eta$ over $\{0.1, 0.25, 0.5, 0.75, 1, 1.5, 2, 4, 10\}$ and trained each setting from scratch under the same evaluation protocol and fixed experimental settings. An Anchor-only reference predicted directly from $\mathbf{z}=\mathbf{z}^{A}$ without relational refinement.

As shown in Fig.~\ref{fig:relational_refinement}(b) and (c), increasing $\eta$ progressively constrained deviation from the patient anchor. Performance peaked at the moderate setting $\eta=1$, which improved balanced accuracy over Anchor only by 5.40 points (95\% CI, 2.18--9.04); positive paired bootstrap intervals were also observed at $\eta=0.5$ and $0.75$, but not consistently at the weakly or strongly anchored extremes.
Patient-level transitions further showed that weak anchoring introduced larger but more harmful corrections, whereas strong anchoring suppressed potentially useful refinement. At $\eta=1$, relational refinement achieved the largest net correction, supporting a balance between resolving errors and preserving already reliable patient-specific evidence.

\subsubsection{Regional Attention Grounding}

\begin{figure}[!ht]
\centering
\includegraphics[width=0.98\textwidth,trim=3bp 8bp 0bp 0bp,clip]{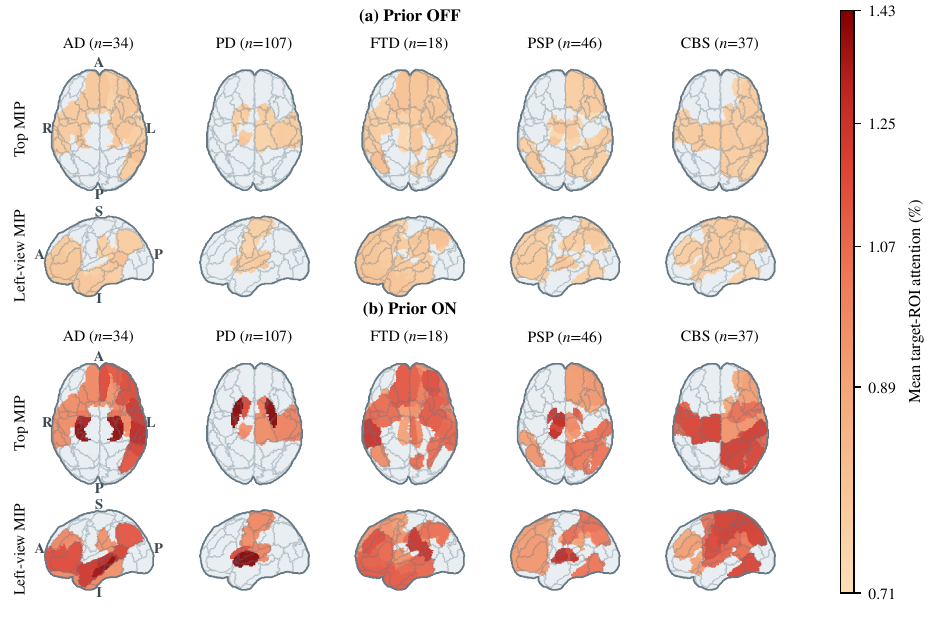}
\caption{Regional attention grounding by structured medical Cards. Mean out-of-fold regional attention is shown with the Card-derived prior (a) disabled and (b) enabled. Colored regions denote prespecified Card targets, with non-target regions faded. Both panels share the same color scale.}
\label{fig:prior_grounding}
\end{figure}

We also compared regional attention with and without the Card-derived prior using the same trained checkpoints. Across the five disease classes, enabling the prior consistently increased attention to prespecified target ROIs (Fig.~\ref{fig:prior_grounding}).
To account for differences in target-set size, we measured target enrichment as mean attention per target ROI relative to non-target ROIs. Enrichment remained near unity without the prior (0.986--1.009) but increased to 1.514--1.655 when enabled, with disease-specific percentile-bootstrap confidence intervals for within-patient changes excluding zero. 
Thus, the Card-derived prior selectively concentrates attention on predefined anatomical targets rather than inducing diffuse redistribution.

\subsubsection{Dose-Dependent Predictive Relevance of Card-Defined Evidence}

\begin{figure}[!htbp]
\centering
\includegraphics[width=\columnwidth]{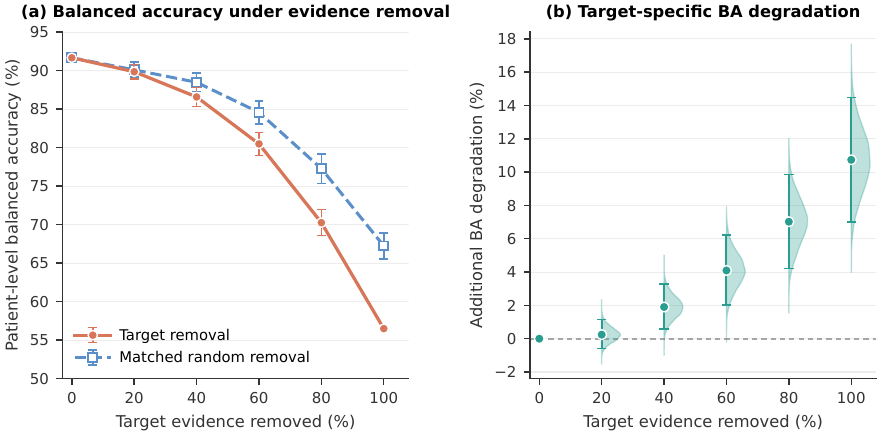}
\caption{Dose-dependent predictive relevance of Card-defined evidence. (a) Patient-level balanced accuracy after progressive Card-target removal and matched random removal of equal numbers of available inputs. (b) Additional target-specific degradation, $\Delta\mathrm{BA}=\mathrm{BA}_{\mathrm{random}}-\mathrm{BA}_{\mathrm{target}}$. Error bars in (a) summarize perturbation variability; points and intervals in (b) show observed differences and 95\% bootstrap confidence intervals.}
\label{fig:target_removal_dose_response}
\end{figure}

Attention grounding alone does not establish that emphasized evidence contributes to prediction. We therefore performed a controlled removal analysis of Card-defined ROI and biomarker targets. Across 331 pooled out-of-fold patients, disease-specific Card targets occupied only $18.01{\pm}5.01\%$ of available evidence slots on average.
Using frozen five-fold checkpoints, we progressively removed 0, 20, 40, 60, 80, or 100\% of each held-out patient's Card-defined targets for the true class. The true label was used only post hoc to define the perturbation set and was never provided during ordinary prediction. Matched-random controls removed identical numbers of available ROI and biomarker inputs. Both perturbations were repeated 100 times with nested removal, while unavailable biomarkers remained masked.

Card-target removal caused a steeper decline in balanced accuracy than matched-random removal, with the separation increasing with removal dose (Fig.~\ref{fig:target_removal_dose_response}). Target-specific degradation was supported from 40\% removal onward, increasing from 1.90 points at 40\% (95\% CI, 0.57--3.29) to 10.74 points at complete removal (95\% CI, 6.99--14.48). At 100\% removal, balanced accuracy fell to 56.50\%, compared with 67.24\% under matched-random removal; macro-F1 showed the same pattern, with an additional 11.78-point degradation (95\% CI, 7.96--15.62).
Together, these analyses show that structured Cards both direct attention toward prespecified medical targets and identify evidence that carries disproportionate predictive information. The increasing degradation with removal dose further suggests that this information is distributed across the Card-defined evidence set rather than confined to a few isolated inputs.

\section{Conclusion}
%这里感觉逻辑上方法，实验结果，又回去讲方法，有点绕，而且分三段太多了。
% Neurodegenerative differential diagnosis requires integrating temporally resolved brain activity, incomplete nonimaging observations, and structured evidence across clinically overlapping disorders. KnowSR addresses this challenge by combining ROI-wise temporal encoding with patient-grounded structured Cards and anchored reconciliation of disease-level and pairwise evidence. Patient-disjoint evaluation across four research databases and 19 baselines demonstrated strong six-class diagnostic performance.

% Ablation and evidence analyses showed that the gains arise from cross-modal interaction, clinically aligned Card semantics, and relational refinement. The latter was particularly informative for the challenging PSP--CBS differential, while regional grounding and controlled evidence removal further showed that Cards direct attention toward predefined anatomical targets with disproportionate predictive relevance.

% By organizing medical knowledge offline and retaining patient-derived representations as the diagnostic anchor, KnowSR provides reproducible knowledge-guided refinement without online retrieval during inference. Together, these results support structured, disease-specific evidence modeling as a principled approach to neurodegenerative differential diagnosis. Future work will evaluate KnowSR in larger independent populations and prospective clinical settings.

%这里再看一下看看表达意思有无偏差，一般conclusion只有一段
\dsyrev{KnowSR combines ROI-wise temporal encoding, patient-grounded structured medical Cards, and anchored reconciliation of disease-level and pairwise evidence for neurodegenerative differential diagnosis. 
Medical knowledge is organized offline, without online retrieval during inference. 
Evaluation on four research databases under patient-disjoint five-fold splits demonstrated improved six-class diagnostic performance over 19 baselines.
Ablation analyses showed contributions from cross-modal interaction, clinically aligned Card semantics, and relational refinement, with the latter particularly benefiting PSP--CBS discrimination. 
Regional attention and evidence-removal analyses further showed that Cards guide attention toward predefined anatomical targets and prioritize evidence relevant to prediction, supporting the value of grounding structured medical knowledge in patient observations.
Future work will evaluate KnowSR in larger independent populations and prospective clinical settings.}
\bibliographystyle{elsarticle-harv}
\bibliography{IEEEabrv,references}

@STRING{IEEE_J_SP         = "{IEEE} Trans. Signal Process."}

@STRING{IEEE_J_MI         = "{IEEE} Trans. Med. Imag."}

@STRING{MICCAI = "Med. Image Comput. Comput. Assist. Interv."}

@String{AAAI = "AAAI"}

@article{jack2024revised,
  author  = {Jack, Jr., Clifford R. and Andrews, J. Scott and Beach, Thomas G. and others},
  title   = {Revised Criteria for Diagnosis and Staging of {Alzheimer}'s Disease: {Alzheimer}'s Association Workgroup},
  journal = {Alzheimers Dement.},
  volume  = {20},
  number  = {8},
  pages   = {5143--5169},
  year    = {2024},
  doi     = {10.1002/alz.13859}
}

@article{postuma2015mds,
  author  = {Postuma, Ronald B. and Berg, Daniela and Stern, Matthew and others},
  title   = {{MDS} Clinical Diagnostic Criteria for {Parkinson}'s Disease},
  journal = {Movement Disorders},
  volume  = {30},
  number  = {12},
  pages   = {1591--1601},
  year    = {2015},
  doi     = {10.1002/mds.26424}
}

@article{rascovsky2011bvftd,
  author  = {Rascovsky, Katya and Hodges, John R. and Knopman, David and others},
  title   = {Sensitivity of Revised Diagnostic Criteria for the Behavioural Variant of Frontotemporal Dementia},
  journal = {Brain},
  volume  = {134},
  number  = {9},
  pages   = {2456--2477},
  year    = {2011},
  doi     = {10.1093/brain/awr179}
}

@article{hoglinger2017psp,
  author  = {H{\"o}glinger, G{\"u}nter U. and Respondek, Gesine and Stamelou, Maria and others},
  title   = {Clinical Diagnosis of Progressive Supranuclear Palsy: The Movement Disorder Society Criteria},
  journal = {Movement Disorders},
  volume  = {32},
  number  = {6},
  pages   = {853--864},
  year    = {2017},
  doi     = {10.1002/mds.26987}
}

@article{armstrong2013cbd,
  author  = {Armstrong, Melissa J. and Litvan, Irene and Lang, Anthony E. and others},
  title   = {Criteria for the Diagnosis of Corticobasal Degeneration},
  journal = {Neurology},
  volume  = {80},
  number  = {5},
  pages   = {496--503},
  year    = {2013},
  doi     = {10.1212/WNL.0b013e31827f0fd1}
}

@article{adni_database,
  author  = {Mueller, Susanne G. and Weiner, Michael W. and Thal, Leon J. and Petersen, Ronald C. and Jack, Clifford R. and Jagust, William and Trojanowski, John Q. and Toga, Arthur W. and Beckett, Laurel},
  title   = {Ways toward an Early Diagnosis in {Alzheimer}'s Disease: The {Alzheimer}'s Disease Neuroimaging Initiative ({ADNI})},
  journal = {Alzheimers Dement.},
  volume  = {1},
  number  = {1},
  pages   = {55--66},
  year    = {2005},
  doi     = {10.1016/j.jalz.2005.06.003}
}

@article{ppmi_database,
  author  = {{Parkinson Progression Marker Initiative}},
  title   = {The {Parkinson} Progression Marker Initiative ({PPMI})},
  journal = {Progress in Neurobiology},
  volume  = {95},
  number  = {4},
  pages   = {629--635},
  year    = {2011},
  doi     = {10.1016/j.pneurobio.2011.09.005}
}

@article{nifd_database,
  author  = {Pankov, Aleksandr and Binney, Richard J. and Staffaroni, Adam M. and Kornak, John and Attygalle, Suneth and Schuff, Norbert and Weiner, Michael W. and Kramer, Joel H. and Dickerson, Bradford C. and Miller, Bruce L. and Rosen, Howard J.},
  title   = {Data-Driven Regions of Interest for Longitudinal Change in Frontotemporal Lobar Degeneration},
  journal = {Neuroimage Clin.},
  volume  = {12},
  pages   = {332--340},
  year    = {2016},
  doi     = {10.1016/j.nicl.2015.08.002}
}

@article{rtni_database,
  author  = {Spina, Salvatore and Brown, Jesse A. and Deng, Jersey and others},
  title   = {Neuropathological Correlates of Structural and Functional Imaging Biomarkers in 4-Repeat Tauopathies},
  journal = {Brain},
  volume  = {142},
  number  = {7},
  pages   = {2068--2081},
  year    = {2019},
  doi     = {10.1093/brain/awz122}
}

@article{kawahara2017brainnetcnn,
  author  = {Kawahara, Jeremy and Brown, Colin J. and Miller, Steven P. and Booth, Brian G. and Chau, Vann and Grunau, Ruth E. and Zwicker, Jill G. and Hamarneh, Ghassan},
  title   = {{BrainNetCNN}: Convolutional Neural Networks for Brain Networks; Towards Predicting Neurodevelopment},
  journal = {NeuroImage},
  volume  = {146},
  pages   = {1038--1049},
  year    = {2017},
  doi     = {10.1016/j.neuroimage.2016.09.046}
}

@article{li2021braingnn,
  author  = {Li, Xiaoxiao and Zhou, Yuan and Dvornek, Nicha C. and Zhang, Muhan and Gao, Siyuan and Zhuang, Juntang and Scheinost, Dustin and Staib, Lawrence H. and Ventola, Pamela and Duncan, James S.},
  title   = {{BrainGNN}: Interpretable Brain Graph Neural Network for {fMRI} Analysis},
  journal = {Medical Image Analysis},
  volume  = {74},
  pages   = {102233},
  year    = {2021},
  doi     = {10.1016/j.media.2021.102233}
}

@article{zhang2023lggnn,
  author  = {Zhang, Hao and Song, Ran and Wang, Liping and Zhang, Lin and Wang, Dawei and Wang, Cong and Zhang, Wei},
  title   = {Classification of Brain Disorders in {rs-fMRI} via Local-to-Global Graph Neural Networks},
  journal = IEEE_J_MI,
  volume  = {42},
  number  = {2},
  pages   = {444--455},
  year    = {2023},
  doi     = {10.1109/TMI.2022.3219260}
}

@inproceedings{kan2022brainnetworktransformer,
  author    = {Kan, Xuan and Dai, Wei and Cui, Hejie and Zhang, Zilong and Guo, Ying and Yang, Carl},
  title     = {Brain Network Transformer},
  booktitle = {Adv. Neural Inf. Process. Syst.},
  volume    = {35},
  pages     = {25586--25599},
  year      = {2022}
}

@inproceedings{caro2023brainlm,
  author    = {Ortega Caro, Josue and de Oliveira Fonseca, Antonio Henrique and Rizvi, Syed A. and Rosati, Matteo and Averill, Christopher and Cross, James L. and Mittal, Prateek and Zappala, Emanuele and Dhodapkar, Rahul Madhav and Abdallah, Chadi and van Dijk, David},
  title     = {{BrainLM}: A Foundation Model for Brain Activity Recordings},
  booktitle = {Int. Conf. Learn. Represent.},
  year      = {2024}
}

@article{qiu2022multimodal,
  author  = {Qiu, Shangran and Miller, Matthew I. and Joshi, Prajakta S. and others},
  title   = {Multimodal Deep Learning for {Alzheimer}'s Disease Dementia Assessment},
  journal = {Nat. Commun.},
  volume  = {13},
  number  = {1},
  pages   = {3404},
  year    = {2022},
  doi     = {10.1038/s41467-022-31037-5}
}

@article{xue2024ai,
  author  = {Xue, Chonghua and Kowshik, Sahana S. and Lteif, Diala and others},
  title   = {{AI}-Based Differential Diagnosis of Dementia Etiologies on Multimodal Data},
  journal = {Nat. Med.},
  volume  = {30},
  number  = {10},
  pages   = {2977--2989},
  year    = {2024},
  doi     = {10.1038/s41591-024-03118-z}
}

@article{golovanevsky2022multimodal,
  author  = {Golovanevsky, Michal and Eickhoff, Carsten and Singh, Ritambhara},
  title   = {Multimodal Attention-Based Deep Learning for {Alzheimer}'s Disease Diagnosis},
  journal = {Journal of the American Medical Informatics Association},
  volume  = {29},
  number  = {12},
  pages   = {2014--2022},
  year    = {2022},
  doi     = {10.1093/jamia/ocac168}
}

@article{yu2024transformer,
  author  = {Yu, Qi and Ma, Qian and Da, Lijuan and Li, Jiahui and Wang, Mengying and Xu, Andi and Li, Zilin and Li, Wenyuan and {{Alzheimer's Disease Neuroimaging Initiative}}},
  title   = {A Transformer-Based Unified Multimodal Framework for {Alzheimer}'s Disease Assessment},
  journal = {Comput. Biol. Med.},
  volume  = {180},
  pages   = {108979},
  year    = {2024},
  doi     = {10.1016/j.compbiomed.2024.108979}
}

@inproceedings{havaei2016hemis,
  author    = {Havaei, Mohammad and Guizard, Nicolas and Chapados, Nicolas and Bengio, Yoshua},
  title     = {{HeMIS}: Hetero-Modal Image Segmentation},
  booktitle = {Medical Image Computing and Computer-Assisted Intervention -- MICCAI 2016},
  series    = {Lecture Notes in Computer Science},
  volume    = {9901},
  pages     = {469--477},
  publisher = {Springer},
  year      = {2016},
  doi       = {10.1007/978-3-319-46723-8_54}
}

@inproceedings{yu2024smart,
  author    = {Yu, Zhihao and Chu, Xu and Jin, Yujie and Wang, Yasha and Zhao, Junfeng},
  title     = {{SMART}: Towards Pre-Trained Missing-Aware Model for Patient Health Status Prediction},
  booktitle = {Adv. Neural Inf. Process. Syst.},
  volume    = {37},
  year      = {2024},
  doi       = {10.52202/079017-2043}
}

@inproceedings{han2024fusemoe,
  author    = {Han, Xing and Nguyen, Huy and Harris, Carl and Ho, Nhat and Saria, Suchi},
  title     = {{FuseMoE}: Mixture-of-Experts Transformers for Fleximodal Fusion},
  booktitle = {Adv. Neural Inf. Process. Syst.},
  volume    = {37},
  year      = {2024},
  doi       = {10.52202/079017-2167}
}

@inproceedings{selvaraju2017gradcam,
  author    = {Selvaraju, Ramprasaath R. and Cogswell, Michael and Das, Abhishek and Vedantam, Ramakrishna and Parikh, Devi and Batra, Dhruv},
  title     = {{Grad-CAM}: Visual Explanations from Deep Networks via Gradient-Based Localization},
  booktitle = {Proc. IEEE Int. Conf. Comput. Vis.},
  pages     = {618--626},
  year      = {2017}
}

@inproceedings{lundberg2017unified,
  author    = {Lundberg, Scott M. and Lee, Su-In},
  title     = {A Unified Approach to Interpreting Model Predictions},
  booktitle = {Adv. Neural Inf. Process. Syst.},
  volume    = {30},
  pages     = {4765--4774},
  year      = {2017}
}

@article{chandak2023primekg,
  author  = {Chandak, Payal and Huang, Kexin and Zitnik, Marinka},
  title   = {Building a Knowledge Graph to Enable Precision Medicine},
  journal = {Sci. Data},
  volume  = {10},
  number  = {1},
  pages   = {67},
  year    = {2023},
  doi     = {10.1038/s41597-023-01960-3}
}

@article{singhal2023clinical,
  author  = {Singhal, Karan and Azizi, Shekoofeh and Tu, Tao and others},
  title   = {Large Language Models Encode Clinical Knowledge},
  journal = {Nature},
  volume  = {620},
  number  = {7972},
  pages   = {172--180},
  year    = {2023},
  doi     = {10.1038/s41586-023-06291-2}
}

@article{thirunavukarasu2023llm,
  author  = {Thirunavukarasu, Arun James and Ting, Darren Shu Jeng and Elangovan, Kabilan and Gutierrez, Laura and Tan, Ting Fang and Ting, Daniel Shu Wei},
  title   = {Large Language Models in Medicine},
  journal = {Nat. Med.},
  volume  = {29},
  number  = {8},
  pages   = {1930--1940},
  year    = {2023},
  doi     = {10.1038/s41591-023-02448-8}
}

@article{Du2026DementiaED,
  title={Dementia Etiology Diagnosis via Collaborative Meta Knowledge Enhancement},
  author={Siyuan Du and Mengxi Chen and Xinyang Jiang and Zilong Wang and Jiangchao Yao and Dongsheng Li and Ya Zhang and Lili Qiu and Yanfeng Wang},
  journal=IEEE_J_MI,
  year={2026},
  volume={PP},
  url={https://api.semanticscholar.org/CorpusID:290623170}
}

@article{cortes1995support,
  author  = {Cortes, Corinna and Vapnik, Vladimir},
  title   = {Support-Vector Networks},
  journal = {Machine Learning},
  volume  = {20},
  number  = {3},
  pages   = {273--297},
  year    = {1995},
  doi     = {10.1007/BF00994018}
}

@article{breiman2001random,
  author  = {Breiman, Leo},
  title   = {Random Forests},
  journal = {Machine Learning},
  volume  = {45},
  number  = {1},
  pages   = {5--32},
  year    = {2001},
  doi     = {10.1023/A:1010933404324}
}

@inproceedings{chen2016xgboost,
  author    = {Chen, Tianqi and Guestrin, Carlos},
  title     = {{XGBoost}: A Scalable Tree Boosting System},
  booktitle = {Proc. 22nd ACM SIGKDD Int. Conf. Knowl. Discov. Data Min.},
  pages     = {785--794},
  year      = {2016},
  doi       = {10.1145/2939672.2939785}
}

@article{schuster1997bidirectional,
  author  = {Schuster, Mike and Paliwal, Kuldip K.},
  title   = {Bidirectional Recurrent Neural Networks},
  journal = IEEE_J_SP,
  volume  = {45},
  number  = {11},
  pages   = {2673--2681},
  year    = {1997},
  doi     = {10.1109/78.650093}
}

@inproceedings{tsai2019multimodal,
  author    = {Tsai, Yao-Hung Hubert and Bai, Shaojie and Liang, Paul Pu and Kolter, J. Zico and Morency, Louis-Philippe and Salakhutdinov, Ruslan},
  title     = {Multimodal Transformer for Unaligned Multimodal Language Sequences},
  booktitle = {Proc. 57th Annu. Meeting Assoc. Comput. Linguist.},
  pages     = {6558--6569},
  year      = {2019},
  doi       = {10.18653/v1/P19-1656}
}

@inproceedings{xu2024contrasformer,
  author    = {Xu, Jiaxing and He, Kai and Lan, Mengcheng and Bian, Qingtian and Li, Wei and Li, Tieying and Ke, Yiping and Qiao, Miao},
  title     = {{Contrasformer}: A Brain Network Contrastive Transformer for Neurodegenerative Condition Identification},
  booktitle = {Proc. 33rd ACM Int. Conf. Inf. Knowl. Manag.},
  pages     = {2671--2681},
  year      = {2024},
  doi       = {10.1145/3627673.3679560}
}

@article{kim2021stagin,
  author  = {Kim, Byung-Hoon and Ye, Jong Chul and Kim, Jae-Jin},
  title   = {Learning Dynamic Graph Representation of Brain Connectome with Spatio-Temporal Attention},
  journal = {Adv. Neural Inf. Process. Syst.},
  volume  = {34},
  pages   = {4314--4327},
  year    = {2021}
}

@inproceedings{xue2025dhgformer,
  author    = {Xue, Rundong and Hu, Hao and Zhang, Zeyu and Han, Xiangmin and Wang, Juan and Gao, Yue and Du, Shaoyi},
  title     = {{DHGFormer}: Dynamic Hierarchical Graph Transformer for Disorder Brain Disease Diagnosis},
  booktitle = {Medical Image Computing and Computer Assisted Intervention -- MICCAI 2025},
  pages     = {268--278},
  year      = {2025},
  publisher = {Springer}
}

@inproceedings{pei2025cagt,
  author    = {Pei, Shengbing and Ma, Jiajun and Lv, Zhao and Zhang, Chao and Guan, Jihong},
  title     = {Community-Aware Graph Transformer for Brain Disorder Identification},
  booktitle = {Proceedings of the Thirty-Fourth International Joint Conference on Artificial Intelligence},
  pages     = {4191--4199},
  year      = {2025},
  doi       = {10.24963/ijcai.2025/467}
}

@inproceedings{yao2024drfuse,
  author    = {Yao, Wenfang and Yin, Kejing and Cheung, William K. and Liu, Jia and Qin, Jing},
  title     = {{DrFuse}: Learning Disentangled Representation for Clinical Multi-Modal Fusion with Missing Modality and Modal Inconsistency},
  booktitle = {Proceedings of the AAAI Conference on Artificial Intelligence},
  volume    = {38},
  number    = {15},
  pages     = {16416--16424},
  year      = {2024},
  doi       = {10.1609/aaai.v38i15.29578}
}

@inproceedings{hayat2022medfuse,
  author    = {Hayat, Nasir and Geras, Krzysztof J. and Shamout, Farah E.},
  title     = {{MedFuse}: Multi-Modal Fusion with Clinical Time-Series Data and Chest X-Ray Images},
  booktitle = {Proceedings of the 7th Machine Learning for Healthcare Conference},
  series    = {Proceedings of Machine Learning Research},
  volume    = {182},
  pages     = {479--503},
  year      = {2022},
  publisher = {PMLR}
}

@article{xu2024contrastpool,
  author  = {Xu, Jiaxing and Bian, Qingtian and Li, Xinhang and Zhang, Aihu and Ke, Yiping and Qiao, Miao and Zhang, Wei and Sim, Wei Khang Jeremy and Gulyas, Balazs},
  title   = {Contrastive Graph Pooling for Explainable Classification of Brain Networks},
  journal = IEEE_J_MI,
  volume  = {43},
  number  = {9},
  pages   = {3292--3305},
  year    = {2024},
  doi     = {10.1109/TMI.2024.3392988}
}

@inproceedings{zhang2025gdl,
  author    = {Zhang, Shengjie and Jiang, Zhuangzhuang and Shen, Xin and Yu, Ziqi and Chen, Xiang and Zhang, Xiao-Yong and Zhou, Yuan},
  title     = {Graph Disentanglement Learning for {fMRI} Analysis: Decoupling Disease, Covariates, and Individual Variability},
  booktitle = {Medical Image Computing and Computer Assisted Intervention -- MICCAI 2025},
  pages     = {352--361},
  year      = {2025},
  publisher = {Springer}
}

@article{cox1958regression,
  author  = {Cox, David R.},
  title   = {The Regression Analysis of Binary Sequences},
  journal = {Journal of the Royal Statistical Society: Series B (Methodological)},
  volume  = {20},
  number  = {2},
  pages   = {215--232},
  year    = {1958},
  doi     = {10.1111/j.2517-6161.1958.tb00292.x}
}

\end{document}